# Large-scale Offshore Wind Farm Electrical Collector System Planning: A Mixed-Integer Linear Programming Approach

Xinwei Shen, *Senior Member, IEEE*, Sunwei Li, Hongke Li

*Abstract*—In this paper, we propose a planning method for large-scale offshore wind farm (OWF) electrical collector system (ECS) based on mixed integer linear programming, in which the sizing and siting of offshore substations and the lines between wind turbines (WTs) are optimized. We found out that the problem is similar to power distribution system planning, where the topological constraints for distribution network expansion planning (DNEP) are applied to guarantee the radiality of ECS topology and accelerate the solving process. Case studies based on an OWF with 63 fixed-locations WTs demonstrate the effectiveness of proposed method, in which the cost of ECS's investment on cables is reduced by 23%, power loss reduced by 44% compared with a conventional design, and the calculation time reduced with the help of the radiality constraint.

## NOMENCLATURE

**Sets and parameters**:

| | |
|---|---|
| $\Psi$ | node set, with $|\Psi|$ nodes |
| $\Psi^{\mathrm{SUB}}/\Psi^{\mathrm{WT}}$ | substation/wind turbine node set |
| $L$ | line set, with $|L|$ lines |
| $L_j$ | lines connected with node $j$ |
| $(i, j)$ | a line between nodes $i$ and $j$ |
| $z_{i,j}$ | Impedance of line $(i, j)$ |
| $\mathbf{S}$ | Bus-line incidence matrix of OWF ECS |
| $g_i^{WT}$ | Wind power generation at node $i$ |
| $c_{i,j}$ | cost of line $(i, j)$, unit: ¥ |
| $\bar{f}_{i,j}$ | the transmission capacity of line $(i, j)$ |
| $\underline{v}/\bar{v}$ | voltage lower/upper limits, unit: p. u. |
| $\eta$ | coefficient of power loss to the whole life-time (planning years) |
| $M$ | Big-M constant |

**Variables**:

| | |
|---|---|
| $x_{i,j}$ | investment decision variable for adding $(i, j)$, $x_{i,j}=1$ if $(i, j)$ is selected, else $x_{i,j}=0$ |
| $\beta_{i,j}^+$ | parent-child relationship of node $i$ and $j$, $\beta_{i,j}^+ =1$ if node $i$ is parent node of node $j$, otherwise $\beta_{i,j}^+ =0$ |
| $f_{i,j}$ | power flow/current in $(i, j)$ |
| $v_i$ | voltage at node $i$, p. u. |
| $g_i^{\mathrm{Cur}}$ | Curtailed wind power at node $i$ |

## I. INTRODUCTION

With rising concerns on carbon emissions and climate change globally, the development of renewable energy in electric power system has been conducted dramatically in the past decades. Offshore wind farm (OWF), with its reducing levelized cost of energy and less requirements on land usage, has become rather competitive in China and all over the world these years. Moreover, for most OWFs, a utilization time of more than 3000 hours per year is to be expected[1], significantly higher than that for on-land sited turbines and, therefore, to a certain extent compensating for the additional costs of offshore plants. In China, the China coastal waters have been identified as suitable for the construction of OWFs thanks to its ambient wind resources, high frequencies of strong wind observations and good wind power stabilities[2]. In recent years, more and more large-scale OWFs have been constructed. For example, in France, it's reported that Saint-Brieuc OWF[3] is a large-scale OWF with 496 MW total generation capacity, in which 62 8 MW wind turbine (WT) are installed, where 90 km 66 kV cables in ECS is utilized to collect all the electric energy from WTs.

In OWFs construction, electrical collector system (ECS) planning and design are one of the major concerns, since the system reliability can be significantly improved by optimized cabling topology, e. g. single/double-sided ring topology[4], while the investment and power loss in ECS can also be reduced by optimal cabling design. In this regard, many scholars have proposed planning methods for ECS in OWF. In [5], a graph-theoretic minimum spanning tree algorithm has been used in the layout of cabling system for OWF, in which the total trenching length is optimized. In [6], three cabling structures for OWF are explored: the string (radial), ring, and multiloop structure, it's found out that multiloop structure increases reliability and proves to be most economic when the failure rate and mean time to repair of cables are relatively high. Some research proposed to model the ECS planning as the well-known capacitated vehicle routing problem (CVRP)[7][8], in which some specific algorithm, e. g. Clarke and

This work is in part supported by National Natural Science Foundation of China (No. 52007123). Dr. X. Shen, Prof. S. Li are with Tsinghua Shenzhen International Graduate School, Tsinghua University, Shenzhen, 518055, China. Hongke Li is with PowerChina Huadong Engineering Corp. Limited. (Corresponding: Xinwei Shen, sxw.tbsi@sz.tsinghua.edu.cn)

Wright savings algorithm[8], could be applied. However, since most of the solving algorithms applied in current research are still heuristic algorithms, e. g. minimum spanning tree[5][7], sweep algorithm/string saving [6], Clarke and Wright savings[7][8] algorithm, NSGA-III and binary PSO[9] etc, the convergence performance is believed to be case-by-case, and optimality gap between their final solutions and global optimal solution cannot be produced (and usually worse than branch & bound/cut[10]). Moreover, to the best of our knowledge, latest progress on simplified power flow model was not presented in aforementioned research, while in most literature AC power flow models are still applied.

On the other hand, distribution network expansion planning (DNEP) problem has been studied by other scholars for decades, in which the mathematical optimization models, especially considering the radiality of the network, are discussed, explored and applied comprehensively, showing its great benefits in reducing distribution system power losses and saving unnecessary investment costs on distribution lines/cables. For instance, it was proved that second-order cone programming and its relaxation[12] are effective for solving optimal power flow problems in radial distribution network, while some other research applied linearization[13] or piecewise linearization[14] technique to simplify DNEP.

In this paper, we propose a mixed-integer linear programming (MILP) model, originally from DNEP, to solve OWF-ECS planning (OWF-ECSP) problem, in which some mature technique in DNEP could also be applied to help produce optimal planning results, especially with lower power loss in ECS, thus improving the economy of OWF construction.

The remaining parts of the paper are structured as follows: Section II discusses and depicts the features of OWF-ECSP and DNEP problems, Section III describe the procedure of proposed OWF-ECSP method and formulates the mathematical model, case studies are given in Section IV, Section V concludes the paper.

## II. COMPARISONS OF OWF-ECSP AND DNEP PROBLEM

The features of OWF-ECSP and DNEP are summarized and compared as below in Table I:

Table I Comparsion between OWF-ECS Planing and DNEP Problem

| | OWF-ECSP | DNEP |
|---|---|---|
| **Objectives** | 1) Investment and operation cost minimization | |
| | 2) Reliability enhancement | |
| **Decision Variables** | Siting and sizing for substations/Cable/feeders | |
| **Constraints** | 1) Voltage limits | |
| | 2) Line/cable capacity limits | |
| | 3) Load balance | |
| **Topology** | Usually radial*, but rings/loops are also applicable** | Usually radial in operation, but rings/loops are also applicable if "N-1" is required |
| Voltage Level (kV) | 33/35/66 | 10/12.6/20/35 |
| Typical nodes' number | 50~100+ | 10~100+ |
| Typical candidate lines' number | 100~1000 | 10~100+ |

*According to current situations in China, when mean time to repair is low.

** According to state-of-the-art research, e. g. [6]~[8]

First of all, the objectives of OWF-ECSP and DNEP are almost the same, since they both aim at optimizing the total investment and operation cost, in which line/cable investment options, power losses reduction in the network and value of lost load/generation should be included. Besides, sometimes the reliability of the planning results is also included, with annual outage hours reduction[15] benefiting from distributed generation and energy storage integration. As for the decision variables, the sizing and siting for substations and lines/cables should be included, in which integer/binary variables are used to denote whether or not this candidate option should be chosen, i.e., $x = 1$ means this option is chosen, otherwise $x = 0$. Regarding the constraints group, they should both include constraints on voltage limits, lines/cables capacity limits and power balance, even though the power flow directions in each problem are reversed. In terms of network topology, radial topology is commonly seen in both OWF-ECSP and DNEP, while ring/loop topology are also alternatives to improve the system reliability, especially when "N-1" criterion in the network needs to be considered in planning stage. Moreover, the voltage levels of both problems are somehow similar, which falls in the range of mid-voltage distribution network, i.e., 10~110 kV. Even their problem scales (number of decision variables) are similar, while OWF-ECSP's typical numbers of load/WT nodes and candidate lines are larger than those of DNEP.

To conclude, we believe that OWF-ECSP and DNEP are similar problems from many perspectives. Therefore, lots of technologies in DNEP are also applicable in OWF-ECSP, including but not limited to: convex relaxation[12], linearization[13][15], piecewise linearization[14] (for optimal power flow), topological constraints.

## III. PROPOSED OWF-ECSP METHOD

The proposed OWF-ECSP method is divided into several steps. Firstly, the inputs are the coordinates of all WTs, assuming that this work has been completed by some engineers in micro-siting[9][11] for the OWF.

In **Step 1**, we include a fuzzy c-means method to locate the offshore substation (OS), since it has been proved to be effective in finding the optimal location for OS in many research[7][8].

In **Step 2**, we propose to list all the potential connection cables between neighborhood WTs and OSs *within a pre-*

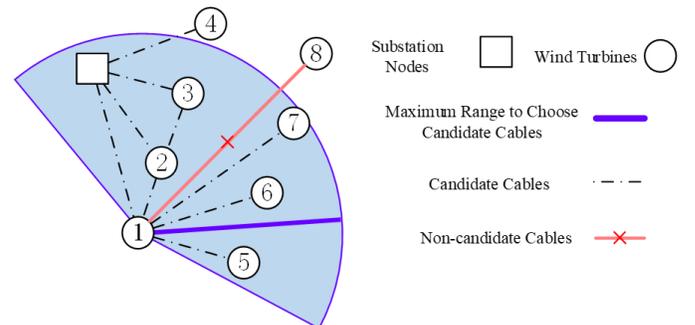

Fig. 1. Step 2: list all the potential connection cables between neighborhood WTs and OSs within a pre-determined maximum range

*determined maximum range*, as illustrated in Fig. 1, which is quite different from other state-of-the-art research. In this way, we assume that the optimal cabling layouts don't require long cables between distanced WTs and OSs, e. g. cables crossing more than 3 WTs, since longer cables also imply larger power losses, and sometimes even not allowed. We do recognize, though, this assumption needs to be further investigated and currently no mathematical proof could be given yet. But similar ideas could be found in early research named as *Granular Tabu Search* for CVRP[16].

For example, in Fig. 1, the potential cables to connect WT 1 is 1-2, 1-5, 1-6, 1-7 and 1-Sub, while 1-4 and 1-8 are non-candidate cables since they are longer than the *maximum range*. The same rules also applied to other WTs to list all the candidate cables. It's noteworthy that, candidate cables for substations can be specially assigned instead of only dependent on the distance.

In **Step 3**, we suggest to list all crossing candidate lines/cables, thus they can be put into the constraints to avoid lines crossings (if necessary). But it's found out that sometimes this step can be saved since line crossings are not happening in the produced planning results.

Then in **Step 4**, we use an MILP model to formulate the OWF-ECSP problem, in which siting and sizing for cables are included, as well as the radiality constraints or other topological constraints to ensure the rationality of final planning result. Then the model is solved by some commercial software, e. g. CPLEX or GUROBI, in **Step 5** and produced the planning result.

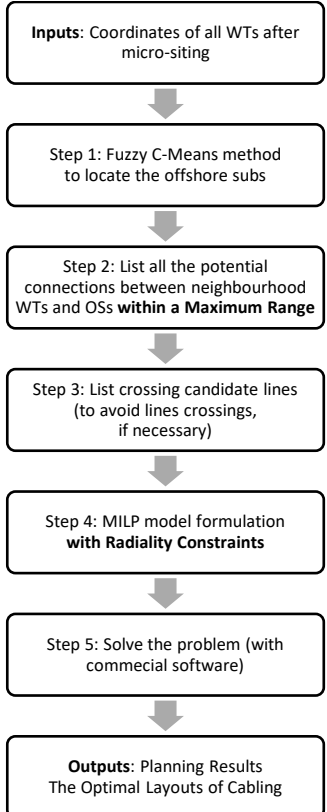

Fig. 2. Proposed OWF-ECSP method's framework

The benefits of the proposed methods are obvious: since we've listed all the candidate cables' options and parameters of all cables are known, with a mixed-integer programming formulation, once we convexify or linearize the optimal power flow problem when the investment decisions are made, i. e. making sub-problem a convex or linear one, the calculation convergence speed can be guaranteed, because we could easily solve the optimization model with branch & bound, branch & cut, benders decomposition and outer approximation. Besides, even though some large-scale MIP problem may also be time-consuming, the optimality of the produced planning results is still known to us, showing how close they are to the global optimizer. Moreover, the model could be implemented and handled by commercial software, saving all the efforts in programming and testing different aforementioned heuristic algorithms.

## IV. OWF-ECSP PROBLEM FORMULATION

For calculation simplicity, in Step 4, we utilized and modified the MILP-based DNEP model stated in [13] and [15], to propose an OWF-ECSP model, in which the optimal power flow model are based on line currents and KCL/KVL laws. The OWF-ECSP model is stated as follows:

$$\min \sum_{(i,j)\in L} c_{i,j} \cdot x_{i,j} + \eta \sum_{(i,j)\in L} r_{i,j} \cdot f_{i,j}^2 + M \cdot \sum_{i\in\Psi^{WT}} g_i^{Cur} \quad (1)$$

s. t.

$$-x_{i,j}\overline{f}_{i,j} \le f_{i,j} \le x_{i,j}\overline{f}_{i,j} \quad \forall(i,j)\in L \quad (2)$$

$$\mathbf{S}\cdot\mathbf{f} = \mathbf{g}-\mathbf{g}^{Cur} \quad (3)$$

$$\left|f_{i,j}\cdot z_{i,j}-(v_i-v_j)\right| \le (1-x_{i,j})\cdot M \quad \forall(i,j)\in L \quad (4)$$

$$\underline{v} \le v_i \le \overline{v} \quad \forall i\in\Psi^{WT} \quad (5)$$

$$0 \le g_i^{Cur} \le g_i \quad \forall i\in\Psi^{WT} \quad (6)$$

$$\sum_{(i,j)\in L} x_{i,j} = \left|\Psi\right|-\left|\Psi^{SUB}\right| \quad (7)$$

$$\beta_{i,j}^+ + \beta_{j,i}^+ = x_{i,j} \quad (8)$$

$$\beta_{i,j}^+ = 1 \quad \forall i\in\Psi^{SUB}, (i,j)\in L \quad (9)$$

$$\sum_{(i,j)\in L_j} \beta_{i,j}^+ = 1 \quad \forall j\in\Psi^{WT} \quad (10)$$

Where the objective function (1) aims at minimizing the cost of constructing ECS cables by summing the cost of each cable $c_{i,j}$ multiplied with its investment decision $x_{i,j}$, and the power losses cost throughout the lifetime of OWF $\eta \sum_{(i,j)\in L} r_{i,j} \cdot f_{i,j}^2$, $\eta$ is the coefficient to transfer power loss at one time slot to the whole lifetime of OWF, i.e. $\eta$=planning years × yearly full load hours (e. g. 3000 hours). The total value of curtailed wind power $M\cdot\sum_{i\in\Psi^{WT}} g_i^{Cur}$ is also included in the objective, but usually not allowed to be not equal to 0 since M is a big number. Constraints (2)-(7) are the operational limits. Constraint (2) means utilization of cable $(i, j)$ is limited by its investment decision $x_{i,j}$, and also denotes the power limits of cables, while constraint (3) denotes the power balance by introducing the node-branch incidence matrix $\mathbf{S}$ and the vectorized power injections of each node ($g_i - g_i^{Cur}$), which comes from imposing KCL. Constraint (4) is a big-M formulation imposing KVL, as eq. (5) of [13] did, while Constraint (5) denotes the voltage limits. Constraint (6) enforces the load not served at each node should be no more than the demand. Constraint (7) implies that the final topology should be $|\Psi^{SUB}|$ trees, and in each tree the substation node is the root node, while the number of cables in operation should be equal to the number of leaf nodes (load nodes). Constraints (8)-(10) are the *spanning tree* constraints resembled from [19] and [20], where (8) guarantees that parent-child relationship of node $i$ and $j$ can be realized if and only if line $(i, j)$ is connected, (9) ensures substation nodes are always the parent node for other nodes, (10) implies only 1 neighbor node can be each WT node's parent node.

One can argue that the accuracy of the MILP model is not enough, but after repeated tests, we found it is a good tradeoff for calculation efficiency while the accuracy in voltages and power flows is also acceptable. In addition, for large-scale OWF's ECS planning problem, it's likely that we could face MIP problem with thousands of integer/binary variables, in which with linear programming as the form of sub-problem would be a good choice to make it possible to converge with acceptable time consumed.

## V. CASE STUDIES

We design 3 cases to show the effectiveness of the proposed method:

**Case 1:** Conventional ECS planning;
**Case 2:** Proposed method without radiality constraint;
**Case 3:** Proposed method with radiality constraint.

In all cases, we solved the OWF-ECS planning model by Gurobi and MATLAB in a laptop with Intel Core i7-1165G7. Since there's not a standard IEEE test case for OWF-ECSP research yet, we resembled the previously-mentioned Saint-Brieuc OWF[3] case with 62 WTs to be a 63 WTs case, with other case conditions the same as its real-world applications[3], e. g. the distance between each row and column. The case conditions are summarized as below:

TABLE II OWF-ECS PLANING: CASE CONDITIONS OF A 63 WTs EXAMPLE

|  | Num of WTs | Capacity of each WT (MW) | Distance between each row (km) | Distance between each column (km) |
|---|---|---|---|---|
| Data | 63 ($7 \times 9$) | 8 | 1 | 1.3 |

The data of case studies, including the coordinates of all WTs as inputs, could be found in [21]. The full load hours in this OWF are assumed to be 2500 hrs., which is lower than that of typical OWFs, usually around 3000-4000.

In Case 1, we assume that 1 fixed location for offshore substation (OS) is predetermined, which locates to the southeast of OWF. As for Case 2 and 3, the location for OS is optimized by proposed method's Step 1, i. e. based on Fuzzy C-Means (FCM) clustering method. After Step 1 and 2, the candidate cables and OS's locations are illustrated below in Fig. 3, where the blue square node is the substation node, yellow ring nodes are all offshore WTs, and dash-dotted lines are candidate cables considered to be built.

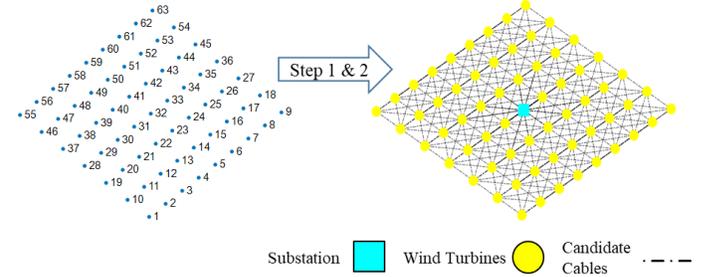

Fig. 3. Candidate cables and OS's locations after Step 1 & 2 in 63-WTs case

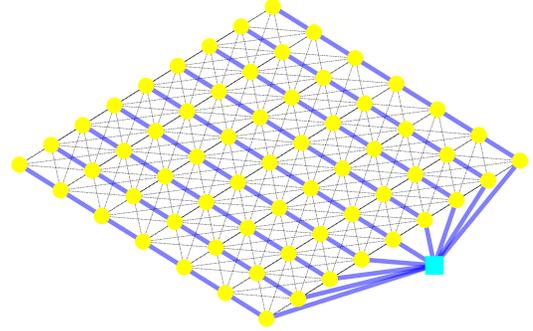

(a) Case 1: Conventional ECS planning scheme

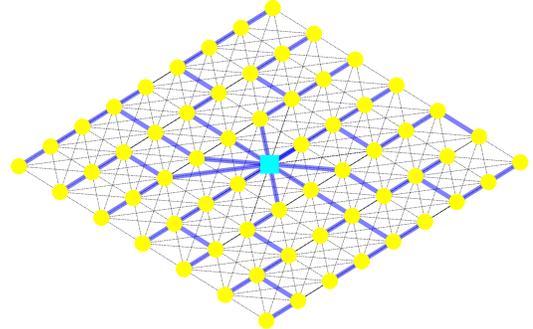

(b) Case 2: Proposed method without radiality constraint

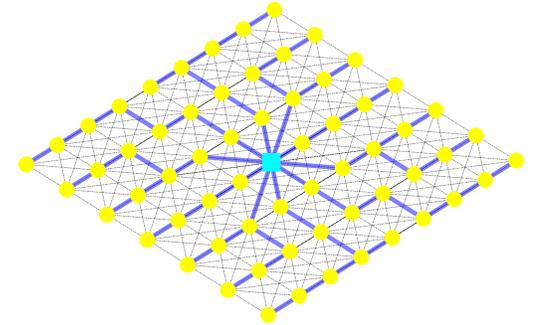

(c) Case 3: Proposed method with radiality constraint

Fig. 4. OWF-ECS planning results: (a) Case 1 (b) Case 2 (c) Case 3

The planning results of different cases are compared in Fig. 4 and Table III. In Fig. 4, the fixed blue lines are newly-built cables, while other dotted dashed lines are lines not chosen. Case 1 showed a conventional planning result, where WTs in each row are connected in a string and delivered to substation.

In Case 2 and 3, OS is located in the center of WTs (with one WT nearby), which definitely reduce the total length of cables, thus reducing the cable investment and power loss. In both Case 2 and Case 3, the network topology is radial, but the details are not the same. Obviously, the topology of Case 3 is symmetric and better than that of Case 2. More details are given in Table III. It could be observed that the radiality constraint is effective in terms of improving the calculation efficiency, as in Case 3 the computation time is significantly reduced compared with that of Case 1 and Case 2.

More importantly, Case 2 and 3 both realize significant total cost reductions compared with Case 1 (26.15%-26.85%), due to the cabling length reduction results in reductions both in cabling investment (23.76%-24.31%) and operation cost (power losses reduced by 42.61%-44.96%). It's also noteworthy that, Case 3 performs better than Case 2 in terms of optimality (optimality gap: 2.66%<6.20%) and calculation efficiency (less time consumption but better solution optimality), which reflects the effectiveness of radiality constraints in proposed MILP model.

Table III OWF-ECS Planing Results Comparion: the 63 WTs Example

| | Case 1 | Case 2 | Case 3 |
|---|---|---|---|
| Inv. Cost (m￥) | 378.01 | 288.18 | 286.12 |
| Ope. Cost (m￥) (power loss in 20 years) | 54.92 | 31.52 | 30.23 |
| Total Cost | 432.93 | 319.70 | 316.35 |
| Cable length (km) | 94.50 | 72.04 | 71.53 |
| $P_{loss}$ rate (%) | 0.256 | 0.147 | 0.141 |
| Cal. Time | 1422.8 | 1090.3 | 772.73 |
| Opt. Gap$^a$ (%) | — | 6.20 | 2.66 |
| Candidate cables | 348 | 354 | 354 |

$^a$The optimality gap, the difference between current best feasible planning solution with known lower bound.

Therefore, it could be concluded that, the proposed OWF-ECSP method outperformed conventional planning scheme in terms of solution optimality. Besides, introducing radiality constraints to further improve its performance is also necessary.

## VI. CONCLUSIONS

In this paper, we compare OWF-ECSP with DNEP problem, find their similarities and propose an MILP-based model for OWF-ECSP correspondingly, which also originated from DNEP area. Unlike conventional research on OWF-ECSP in which heuristic algorithms are usually applied, we propose to list all the candidate cabling choices in advance, thus in proposed model, the optimality of planning result and convergence speed could be guaranteed. With proposed method, case studies showed a significant improvement compared with conventional planning scheme in which offshore substation is predetermined but not optimized, while adding radiality constraint also further improve the model's performance.

It should be noted that, further comparisons between proposed MILP formulation and classical heuristic/metaheuristic algorithms for CVRP are still necessary, especially in terms of solutions' optimality and calculation efficiency. Future research should include SOCP-relaxation-based OWF-ECSP model, as well as some other techniques in the field of DNEP, e. g. piecewise linearization for OWF-ECSP also worth further exploration. Also, the reliability issues can be included in future research, to improve the utilization hours of OWF when cable outage happens.

## REFERENCES

[1] Morthorst, P. E. , and L. Kitzing . "Economics of building and operating offshore wind farms." *Offshore Wind Farms*, pp. 9-27, 2016.

[2] Liu Y, Li S, Chan P W, et al. "On the failure probability of offshore wind turbines in the China coastal waters due to typhoons: a case study using the OC4-DeepCwind semisubmersible". *IEEE Transactions on Sustainable Energy*, vol. 10, no. 2, pp. 522-532, 2018.

[3] Iberdrola, "Saint-Brieuc: Iberdrola's first large-scale offshore wind power project in Brittany", [EB/OL], https://www.iberdrola.com/about-us/lines-business/flagship-projects/saint-brieuc-offshore-wind-farm.

[4] Anaya-Lara O, Tande J O, Uhlen K, et al. "Chapter 6 Offshore Wind Farm Tech. and Electrical Design, Offshore Wind Energy Technology". John Wiley & Sons, 2018.

[5] Dutta, S., & Overbye, T. J. Optimal wind farm collector system topology design considering total trenching length. *IEEE Transactions on Sustainable Energy*, vol. *3*, no. 3, pp 339-348, 2012.

[6] Gong, X., Kuenzel, S., & Pal, B. C.. "Optimal wind farm cabling". *IEEE Transactions on Sustainable Energy*, vol. *9, no.* 3, pp. 1126-1136, 2017.

[7] Zuo, T., Zhang, Y., Meng, K., Tong, Z., Dong, Z. Y., & Fu, Y. "Collector System Topology Design for Offshore Wind Farm's Repowering and Expansion". *IEEE Trans. on Sustainable Energy*, vol. *12*, no. 2, pp. 847-859, 2020.

[8] Zuo, T., Zhang, Y., Meng, K., Tong, Z., Dong, Z. Y., & Fu, Y.. "A Two-Layer Hybrid Optimization Approach for Large-Scale Offshore Wind Farm Collector System Planning". *IEEE Trans. on Industrial Informatics*, accepted, 2021.

[9] Tao, S., Xu, Q., Feijóo, A., & Zheng, G.. "Joint Optimization of Wind Turbine Micrositing and Cabling in an Offshore Wind Farm". *IEEE Transactions on Smart Grid*, vol. 12, no. 1, 834-844, Jan 2021.

[10] G. Laporte, F. Semet, "Chapter 5. Classical Heuristics for the Capacitated VRP", *The Vehicle Routing Problems*, pp. 109-126, 2002.

[11] González, J. S., García, Á. L. T., Payán, M. B., Santos, J. R., & Rodríguez, Á. G. G.. "Optimal wind-turbine micro-siting of offshore wind farms: A grid-like layout approach". *Applied energy*, vol. 200, pp. 28-38, 2017.

[12] Gan L., Li N., Topcu U., & Low S. H. "Exact convex relaxation of optimal power flow in radial networks." *IEEE Transactions on Automatic Control*, vol. 60, no. 1, pp. 72-87, 2015.

[13] Haffner S., et al. "Multistage model for distribution expansion planning with distributed generation—Part I: Problem formulation. *IEEE Transactions on Power Delivery* vol. 23, no. 2, pp. 915-923, 2008.

[14] Zare, A., Chung, C. Y., Zhan, J., & Faried, S. O. A distributionally robust chance-constrained MILP model for multistage distribution system planning with uncertain renewables and loads. *IEEE Transactions on Power Systems*, vol. 33, no.5, pp. 5248-5262, 2018.

[15] Shen, X., Shahidehpour, M., Han, Y., Zhu, S., & Zheng, J.. "Expansion planning of active distribution networks with centralized and distributed energy storage systems". *IEEE Transactions on Sustainable Energy*, vol. 8, no. 1, 126-134, 2017.

[16] Toth, Paolo, Vigo, & Daniele. "The granular tabu search and its application to the vehicle-routing problem". *INFORMS Journal on Computing*, 2003.

[17] Tennet, DNV GL, "66 kV Systems for Offshore Wind Farms- NL Offshore Wind Farm Transmission Systems", [EB/OL], 2015.

[18] Cui D., Jiang C., Shi Z. et al. "Research on Array Submarine Cables in Large Offshore Wind Farm" (in Chinese), *Southern Energy Construction*, vol. 7, no. 2, pp. 98-102, 2020.

[19] J. A. Taylor and F. S. Hover, "Convex models of distribution system reconfiguration," *IEEE Trans. Power Syst.*, vol. 27, no. 3, pp. 1407–1413, Aug. 2012.

[20] M. Lavorato, J. F. Franco, M. J. Rider, and R. Romero, "Imposing radiality constraints in distribution system optimization problems," *IEEE Trans. Power Syst.*, vol. 27, no. 1, pp. 172–180, Feb. 2012.

[21] Data of 63-WTs OWF in Case Studies. 2021 [Online] Available: https://share.weiyun.com/kch4uSZM .